\documentclass[twocolumn,showpacs,preprintnumbers,amsmath,amssymb]{revtex4}

\usepackage{graphicx}% Include figure files
\usepackage{dcolumn}% Align table columns on decimal point
\usepackage{bm}% bold math

\begin{document}
\title{Integrating static and dynamic information for routing traffic}
\author{Wen-Xu Wang$^{1}$}
\author{Chuan-Yang Yin$^{1}$}
\author{Gang Yan$^{2}$}
\author{Bing-Hong Wang$^{1}$}
\email{bhwang@ustc.edu.cn}

\affiliation{%
$^{1}$Nonlinear Science Center and Department of Modern Physics,
University of Science and Technology of China, Hefei, 230026, PR
China  \\
$^{2}$Department of Electronic Science and Technology, University
of Science and Technology of China, Hefei, 230026, PR China
}%

\date{\today}

\begin{abstract}
The efficiency of traffic routing on complex networks can be
reflected by two key measurements i.e. the system capacity and the
average data packets travel time. In this paper, we propose a
mixing routing strategy by integrating local static and dynamic
information for enhancing the efficiency of traffic on scale-free
networks. The strategy is governed by a single parameter.
Simulation results show that there exists a optimal parameter
value by considering both maximizing the network capacity and
reducing the packet travel time. Comparing with the strategy by
adopting exclusive local static information, the new strategy
shows its advantages in improving the efficiency of the system.
The detailed analysis of the mixing strategy is provided. This
work suggests that how to effectively utilize the larger degree
nodes plays the key role in the scale-free traffic systems.
\end{abstract}


\maketitle
\section{Introduction}
Communication networks such as the Internet, World-Wide-Web and
pear-to-pear networks play an important role in modern society.
Dynamical properties of these systems have attracted tremendous
interests and devotion among not only engineering but also physics
communities
\cite{Internet1,Internet2,direct,Hierarchical,Cost,Onset,Optimal2,Trans,ACM,overload}.
The ultimate goal of studying these large communication networks
is to control the increasing traffic congestion and improve the
efficiency of information transportation. Many recent studies have
focused on the efficiency improvement of communication networks
which is usually considered from two aspects: one is modifying the
underlying network structure \cite{Optimal1,Selfadapt,CDcongest}
and the other is developing better routing strategies
\cite{Tadic,Tadic2,Improved,efficient1,efficient2,efficient3}.
Comparing with the former, the latter is preferable with respect
to the high cost of changing underlying structure. In traffic
systems, the structure of the underlying network play an
significant role in the traffic dynamics. In order to develop
practical routing strategies, understanding the effect of network
on the traffic dynamics is a central problem.

Since the surprising discovery of scale-free property of real
world networks by Barab\'si and Albert \cite{BA,Review1}, it is
worthwhile to investigate the traffic dynamics on the scale-free
networks instead of random and regular networks. How the traffic
dynamics are influenced by many kinds of structures, such as
Web-graph \cite{Tadic,Tadic2}, hierarchical trees
\cite{Hierarchical} and Barab\'asi-Albert network \cite{Onset},
has been extensively investigated. A variety of empirically
observed dynamical behaviors have been reproduced by such traffic
models, including $1/f$-like noise of load series, phase
transition from free flow state to congestion, power-law scaling
correlation between flux and the relevant variance and cascading
\cite{Tadic,Tadic2,direct,Netdynamic1,Netdynamic2,Cascade1,Cascade2,Cascade3}.
Moreover, some previous work pointed out that traffic processes
taking place on the networks do also remarkably affect the
evolution of the underlying structure \cite{BBV,WWX}. To model
traffic dynamics on networks, especially for the Internet and WWW,
data-packet generating rate together with their randomly selected
source and destinations are in introduced by previous work
\cite{Tadic3}. Some models assume that packets are routed along
shortest paths from origins to destinations
\cite{Hierarchical,Onset}. However, due to the difficulty in
searching and storing shortest paths between any pair of nodes of
large networks, the routing strategies based on local topological
information have been proposed for better mimicking real traffic
systems and for more widely potential application, such as
Pear-to-Pear networks \cite{Tadic,Tadic2,efficient1}.

Among previous studies, the efficiency of the communication
networks has received much attention for its practical importance
\cite{Hierarchical,Onset,CDcongest,Improved,efficient1,efficient2,efficient3}.
The efficiency of the traffic system can be reflected by capacity
and communication velocity of the system. The capacity is measured
by the onset of phase transition from free flow to congestion,
which has been investigated in several works
\cite{Hierarchical,Onset,efficient1,efficient2,efficient3}. The
phase transition points can be denoted by the maximal packet
generating rate, under which the system is in the free flow state
and no congestion occurs, so that the system can maintain it
normal and efficient functioning. The communication velocity is
measured by the average packet travel time. We note that these two
quantities are not equivalent for estimating the efficiency, but
even contradictory to each other. Take scale-free networks for
example. Scale-free networks possess shorter average path length
in contrast with random and regular networks, which is attributed
to the existence of hub nodes. Thus, data packets can transmit
much faster in the scale-free networks. However, suppose that too
much data flow passes through those hub nodes, it will lead to the
congestion at those nodes and decrease the network capacity.
Aiming to solve the conflict between the network capacity and the
communication speed, we propose a new routing strategy adopting in
the scale-free networks based on local static and dynamic
information, which are the local structure and traffic flux,
respectively. Comparing with the strategy based on exclusive local
static information \cite{efficient1}, both the capacity of the
networks and the average packet travel time are considerably
improved by adopting the mixing strategy. The effects of the new
strategy on the efficiency of the scale-free traffic system are
discussed in detail. The introduced strategy may has potential
application in pear-to-pear networks.

The paper is organized as follows. In the following section, we
describe the model rules and the relevant definitions in detail.
In Sec. III we demonstrate the simulation results and the
discussion. In the last section, the present work is concluded.

\section{The model and definitions}
Barab\'asi-Albert model is the simplest and a well-known model
which can generate networks with power-law degree distribution
$P(k) \thicksim k^{-\gamma}$, where $\gamma=3$. Without losing
generality we construct the network structure by following the
same method used in Ref. \cite{BA}: Starting from $m_0$ fully
connected nodes, a new node with $m_0$ edges is added to the
existing graph at each time step according to preferential
attachment, i.e., the probability $\Pi_i$ of being connected to
the existing node $i$ is proportional to the degree $k_i$. Then we
model the traffic of packets on the given graph. At each time
step, there are $R$ packets generated in the system, with randomly
selected sources and destinations. We treat all the nodes as both
hosts and routers \cite{Tadic,Tadic2,Onset} and assume that each
node can deliver at most $C$ packets per time step towards their
destinations. All the nodes perform a parallel local search among
their immediate neighbors. If a packet's destination is found
within the searched area of node $l$, i.e. the immediate neighbors
of $l$, the packets will be delivered from $l$ directly to its
target and then removed from the system. Otherwise, the
probability of a neighbor node $i$, to which the packet will be
delivered is as follows:
\begin{equation}
P_{l \rightarrow i}=\frac{k_i(n_i+1)^{\beta}}{\sum_j
k_j(n_j+1)^{\beta}},
\end{equation}
where, the sum runs over the immediate neighbors of the node $l$.
$k_i$ is the degree of node $i$ and $n_i$ is the number of packets
in the queue of $i$. $\beta$ is an introduced tunable parameter.
$k_i$ and $n_i$ are the so-called static and dynamic information,
respectively. Adding one to $n_i$ is to guarantee the nodes
without packets have probability to receive packets. During the
evolution of the system, FIFO (first-in first-out) rule is applied
and each packet has no memory of previous steps. Under the control
of the routing strategy, all packets perform a biased random-walk
like move. All the simulations are performed with choosing $C=5$.

To implement our strategy, each node should know the traffic load
of its neighbors, which can be implemented by using the
\textit{keep-alive} messages that routers (nodes) continuously
exchange real-time information with their peers \cite{Improved}.
However, taking into account the information transmission cost,
the exchanged information may be updated every few second between
neighbors. Therefore, we study the effect of transmission delay on
the traffic dynamics. The delay in our model is defined as the
number of time steps (period) of receiving updating information
from neighbors.

\begin{figure}
\scalebox{0.80}[0.80]{\includegraphics{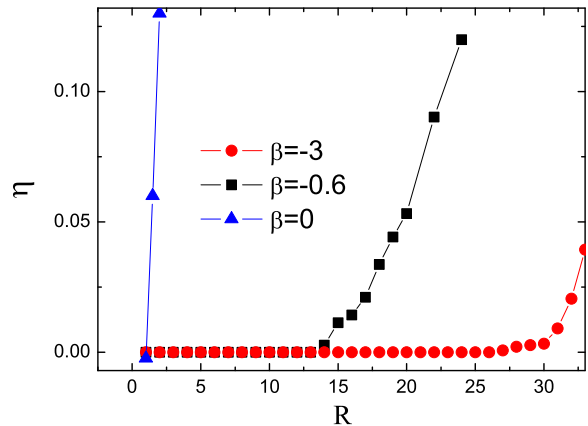}}
\caption{\label{fig:epsart} (color online). The order parameter
$\eta$ as a function of generating rate $R$ for different value of
parameter $\beta$. Other parameters are $delay=0$, $C=5$ and
$N=1000$.}
\end{figure}

In order to characterize the network capacity, we use the order
parameter presented in Ref. \cite{Hierarchical}:
\begin{equation}
\eta(R)=\lim_{t\rightarrow \infty}\frac{\langle\Delta
Load\rangle}{R\cdot\Delta t},
\end{equation}
where $Load(t)$ is defined as the number of packets within the
network at time $t$. $\Delta Load=Load(t+\Delta t)-Load(t)$ with
$\langle\cdots\rangle$ indicates average over time windows of
width $\Delta t$. The order parameter represents the ratio between
the outflow and the inflow of packets calculated over long enough
period. In the free flow state, due to the balance of created and
removed packets, the load does not depend on time, which brings a
steady state. Thus when time tends to be unlimited, $\eta$ is
about zero. Otherwise, when $R$ exceeds a critical value $R_c$,
the packets will continuously pile up within the network, which
destroys the stead state. Thus, the quantities of packets within
the system will be a function of time, which makes $\eta$ a
constant more than zero. Therefore, a sudden increment of $\eta$
from zero to nonzero characterizes the onset of the phase
transition from free flow state to congestion, and the network
capacity can be measured by the maximal generating rate $R_c$ at
the phase transition point.

\begin{figure}
\scalebox{0.80}[0.80]{\includegraphics{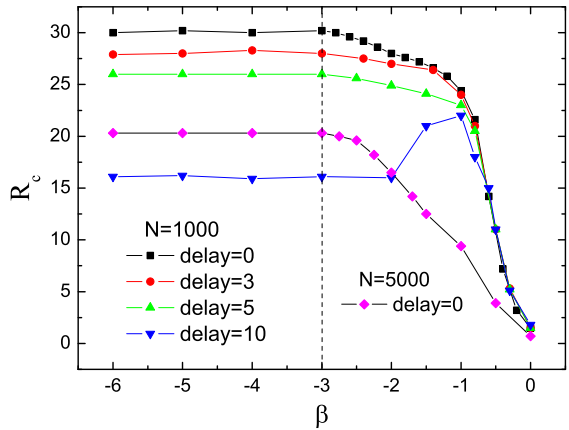}}
\caption{\label{fig:epsart} (color online). The network capacity
$R_c$ versus parameter $\beta$ for different time delay and for
different network size $N$. Other parameter is $C=5$.}
\end{figure}

\section{simulation results}
As mentioned above, the efficiency of the system is reflected by
both the network capacity and the communication velocity. We first
investigate the order parameter $\eta$ as a function of generating
rate $R$ for different model parameter $\beta$, as shown in Fig.
1. One can find that for each $\beta$, when $R$ is less than a
specific value $R_c$, $\eta$ is zero; it suddenly increases when
$R$ is slightly larger than $R_c$. Moreover, in this figure,
different $\beta$ corresponds to different $R_c$, thus we
investigate the network capacity $R_c$ depending on $\beta$ for
finding the optimal value of parameter $\beta$. Figure 2 shows
that in the case of no time delay, the network capacity is
considerably enhanced by reducing $\beta$, and when $\beta$ is
less than a specific value, approximately $-3$, the capacity
reaches an upper limit. The dynamic information $n_i$ represents
the amount of traffic flux of node $i$. The effect of decreasing
$\beta$ is to allow packets to circumvent the nodes with heavier
traffic burden and alleviate the congestion on those nodes. While
for the case of only adopting local topology information (static
strategy) \cite{efficient1,note}, the maximal network capacity is
$23$ with choosing $C=5$. Therefore, the higher maximal capacity
by adopting the new strategy indicates that the dynamic
information is a better reflection of congestion than the static
one. Moreover, the capacity with time delay are also studied, as
shown in Fig. 2. When the delay is not long, $R_c$ is slightly
reduced as increasing the delay, and the onset of the upper limit
is still at $\beta=-3$. However, for the long delay, such as
$delay=10$, it has remarkable influence on the network capacity.
There exists a maximal value of $R_c$ as the point of $\beta=-1$
instead of reaching upper limit and the network capacity is
reduced. The feedback information with long period cannot well
reflect the real circumstance of the neighbor nodes, which leads
to the instabilities in the system, so that the capacity
decreases. Furthermore, we perform simulations with larger network
size, $N=5000$, as exhibited in Fig. 2. The curve of $R_c$ vs
$\beta$ displays the same tendency with the cases of $N=1000$. It
is the longer average shortest path length that results in the
decrease of network capacity comparing with the cases of $N=1000$.

\begin{figure}
\scalebox{0.80}[0.80]{\includegraphics{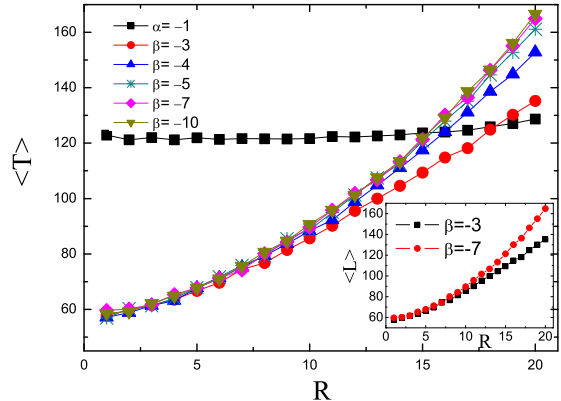}}
\caption{\label{fig:epsart} (color online). Mean packets travel time
$\langle T \rangle$ versus $R$ for static and the new strategy,
respectively. $\alpha=-1$ corresponds to the optimal value of static
strategy. $\beta$ is the parameter of the mixing strategy. The inset
is the average distances travelled by packets $\langle L \rangle$ as
a function of $R$ for different $\beta$. The network size $N=1000$,
node delivering ability $C=5$.}
\end{figure}

The communication velocity of the system can be estimated by the
mean travel time of the packets from their origins to destinations
over a long period. The mean travel time $\langle T \rangle$ vs
creating rate $R$ for different parameter $\beta$ are demonstrated
in Fig. 3. We also compare the behavior of $\langle T \rangle$ by
adopting static strategy \cite{note} with adopting the new one
upon the identical network structure, where $\alpha=-1$
corresponds to the optimal parameter value of the static strategy
as shown in Fig. 3. All the simulations are performed within the
steady state, in which $\langle T \rangle$ is independent of time
step. While if the system enters the jammed state, $\langle T
\rangle$ will increase as time grows, ultimately, it will be prone
to be unlimited due to packets' continuous accumulation in the
system. By adopting the static strategy, $\langle T \rangle$ is
approximately independent of $R$, which is because that the static
routing algorithm is based on exclusive topological information.
Although the static strategy strongly improves the network
capacity, it ignores the importance of hub nodes i.e. greatly
reducing the diameter of the network. In contrast with the static
strategy, the new strategy by integrating the local static and
dynamic information can not only considerably enhance the network
capacity but also make the hub nodes efficiently utilized. One can
see in Fig. 3, when $R$ is not too large, $\langle T \rangle$ by
adopting the new strategy for all $\beta$ is much shorter than
that by adopting static strategy. The advantages of using the new
strategy can be explained from Eq. 1. For very small $R$, few
packets accumulate in the queue, thus $P_i\sim k_i$, which is
consistent with the search algorithm proposed in Ref.
\cite{BJKim}. The high efficiency of this algorithm for searching
target nodes has been demonstrated by numerical simulations
\cite{BJKim}. Hence, it is the shorter average distances travelled
by packets $\langle L \rangle$ that induces the shorter $\langle T
\rangle$ in the free flow state, which can be seen in the inset of
Fig. 3, where $\langle L \rangle$ is nearly the same with $\langle
T \rangle$ for identical value of $\beta$. When increasing $R$,
packets start to accumulate on the large degree nodes, the new
strategy can automatically decrease the probability of packets
delivered to those hub nodes according to the dynamic information.
Then, when $R$ approaches to the $R_c$, packets are routed by the
mixing strategy to circumvent those hub nodes, which become the
bottleneck of the system. Therefore, near the phase transition
point, $\langle T \rangle$ shows the same value by adopting two
different strategies. Combing the results that $\beta=-3$ is not
only the onset of upper limit of the network capacity but also
corresponds to the shortest $\langle T \rangle$ in the case of
maximal network capacity, we can conclude that $\beta=-3$ is the
optimal choice.

\begin{figure}
\scalebox{0.80}[0.80]{\includegraphics{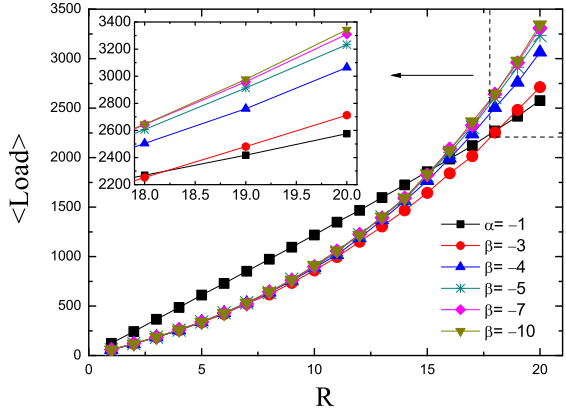}}
\caption{\label{fig:epsart} (color online). Traffic load versus $R$
for static and the new strategy, respectively. $\alpha=-1$
corresponds to the optimal value of static strategy. $\beta$ is the
tunable parameter of the new strategy. The network size $N=1000$,
$C=5$.}
\end{figure}

We further investigate the behavior of traffic load influenced by
the routing strategy. Fig. 3 displays the average load $\langle
Load \rangle$ as a function of $R$ for two strategies with
different parameters. For static strategy with the optimal
parameter, Load is a linear function of $R$. As to the new
strategy, for a wide range of $R$, the load adopting the new
strategy is lower than that by adopting the static one. When $R$
approaches the critical value $R_c$ ,the load with the new
strategy turns to be larger. We also observe that by choosing
$\beta=-3$, the system affords the lowest traffic load among the
whole range of $R$, which also demonstrates that $\beta=-3$ is the
optimal choice. Actually, there exists some relationship between
mean packets travel time and average load. According to the
Little's law \cite{Little} in the queueing theory, one can easily
obtain $<\langle Load \rangle=R\cdot\langle T \rangle$. Note that
this result is only satisfied in the steady state due to the
balance between created and removed nodes.

To give a detailed insight into the effect of the new strategy, we
investigate the queue length of a node $n_k$ as a function of its
degree $k$ with selecting the optimal parameter for different $R$.
The queue length of a node is defined as the number of packets in
the queue of that node. The results are shown in Fig. 4. One can
see that when $R$ is not large, $n_k$ vs $k$ shows power law
properties and the slope for different $R$ is the same. These
behaviors are attributed to the domination of static information.
In Eq. 1, small $R$ leads to the small $n_i$, and the forwarding
probability is mainly determined by the node degree i.e. static
information. Therefore, $n_k$ versus $k$ demonstrates universal
scaling property \cite{efficient1}. For medium value of $R$, such
as $R=20$, the new strategy mainly affects traffic load on the
large degree nodes. The strategy according to Eq.1 allows packets
to circumvent the large degree nodes which bear heavier traffic
burden. When $R$ approaches to the phase transition point $R_c$,
we can see in Fig. 4 that traffic burden on all different degree
nodes is almost the same. This average effect results in the
maximal network capacity in the case of identical node delivering
capacity.

\begin{figure}
\scalebox{0.8}[0.8]{\includegraphics{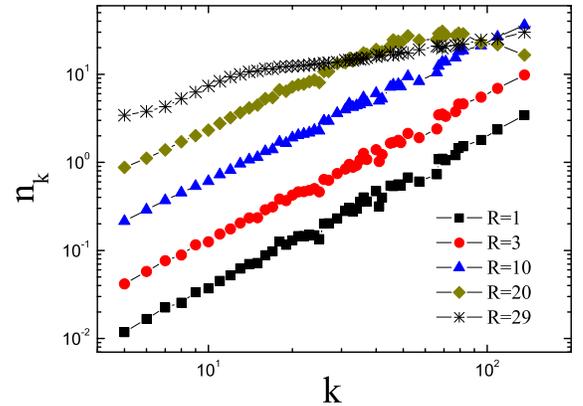}}
\caption{\label{fig:epsart} (color online). The queue length of the
nodes as a function of their degree with $\beta=-3$ for different
$R$. The network size $N=1000$, $C=5$.}
\end{figure}

\section{conclusion}
We have proposed a new routing strategy by integrating local
static and dynamic information. The advantages of this strategy
for delivering data packets on scale-free networks have been
demonstrated from two aspects of network capacity and mean packets
travel time. The short mean packets travel time is mainly due to
the sufficient use of hub nodes. The large network capacity is
caused by the utilization of dynamic information which reflects
the traffic burden on nodes. The present study indicates that the
large degree nodes play an important role in packets delivery. The
packets can find their targets with higher probability if they
pass by the large degree nodes, which results in shorter average
travel time by packets. However, the large degree nodes are also
easily congested if large amount of packets are prone to pass
through them. The introduced strategy can make the large degree
nodes fully used when packet creating rate is low, and also allow
packets to bypass those nodes when they afford heavy traffic
burden. Thus the system's efficiency is greatly improved.

In addition, we note that the new strategy should not be hard for
implementation. The local static i.e. topology information can be
easily acquired and stored in each router. The local dynamic
information could be obtained by using the \textit{keep-alive}
messages that router continuously exchange with their peers
\cite{Improved}. The strategy may has potential application in
pear-to-pear networks.

\section{acknowledgements}
The authors wish to thank Na-Fang Chu for her valuable comments
and suggestions. This work is funded by NNSFC under Grants No.
10472116, 70271070 and 70471033, and by the Specialized Research
Fund for the Doctoral Program of Higher Education (SRFDP
No.20020358009).


\begin{thebibliography}{ref1}
\bibitem{note} Static strategy: the probability of delivering a packet from node
$l$ to node $i$ is $P_i=k_i^{\alpha}/\sum_j k_j^{\alpha}$, where
$k_i$ is the degree of node $i$ and $\alpha$ is a tunable
parameter. The sum runs over the neighbors of node $l$
\cite{efficient1}.
\end{thebibliography}
\end{document}